\documentclass[twocolumn]{aastex63}
\usepackage{amsmath, amsthm,amssymb}
\usepackage{epstopdf}
\usepackage{multirow}
\usepackage{url}

\newcommand{\coldclassical}{{\it cold classical}}
\newcommand{\coldclassicals}{{\it cold classicals}}
\newcommand{\bluebinaries}{{\it blue binaries}}

\begin{document}

\title{Col-OSSOS: The Distinct Colour Distribution of Single and Binary Cold Classical KBOs}

\correspondingauthor{Wesley C. Fraser}
\email{wesley.fraser@nrc-cnrc.gc.ca}
\author{Wesley C. Fraser}
\affiliation{Herzberg Astronomy and Astrophysics Research Centre, National Research Council, 5071 W. Saanich Rd. Victoria, BC V9E 2E7}
\author{Susan D. Benecchi}
\affiliation{Planetary Science Institute, 1700 East Fort Lowell, Suite 106, Tucson, AZ 85719, USA}
\author{JJ Kavelaars}
\affiliation{Herzberg Astronomy and Astrophysics Research Centre, National Research Council, 5071 W. Saanich Rd. Victoria, BC V9E 2E7}
\author{Micha\"{e}l Marsset}
\affiliation{Department of Earth, Atmospheric and Planetary Sciences, MIT, 77 Massachusetts Avenue, Cambridge, MA 02139, USA}
\author{Rosemary E. Pike}
\affiliation{Smithsonian Astrophysical Observatory; 60 Garden Street, Cambridge, MA, 02138, USA}
\author{Michele T. Bannister}
\affiliation{School of Physical and Chemical Sciences — Te Kura Mat\~{u}, University of Canterbury, Private Bag 4800, Christchurch 8140, New Zealand}
\author{Megan E. Schwamb}
\affiliation{Astrophysics Research Centre, Queen’s University Belfast, Belfast BT7 1NN, UK}
\author{Kathryn Volk}
\affiliation{Lunar and Planetary Laboratory, The University of Arizona, 1629 E. University Boulevard, Tucson, AZ 85721, USA}
\author{David Nesvorny}
\affiliation{Department of Space Studies, Southwest Research Institute, 1050 Walnut St., Suite 300, Boulder, CO 80302, USA}
\author{Mike Alexandersen}
\affiliation{Smithsonian Astrophysical Observatory; 60 Garden Street, Cambridge, MA, 02138, USA}
\author{Ying-Tung Chen}
\affiliation{Institute of Astronomy and Astrophysics, Academia Sinica, Taipei 10617, Taiwan}
\author{Stephen Gwyn}
\affiliation{Herzberg Astronomy and Astrophysics Research Centre, National Research Council, 5071 W. Saanich Rd. Victoria, BC V9E 2E7}
\author{Matthew J. Lehner}
\affiliation{Institute of Astronomy and Astrophysics, Academia Sinica, Taipei 10617, Taiwan}
\affiliation{ Department of Physics and Astronomy, University of Pennsylvania, 209 South 33rd Street, Philadelphia, PA 19104, USA}
\affiliation{Harvard-Smithsonian Center for Astrophysics, 60 Garden Street, Cambridge, MA 02138, USA}
\author{Shiang-Yu Wang}
\affiliation{Institute of Astronomy and Astrophysics, Academia Sinica, Taipei 10617, Taiwan}
\date{} 

\begin{abstract}
The cold classical Kuiper Belt Objects (KBOs) possess a high, $\gtrsim30\%$ binary fraction. Widely separated and dynamically fragile, these binary systems have been useful in tracing the origins of KBOs. A new class of binaries was recently identified by their colours. The so-called blue binaries are unanimously members of the less red compositional class, and exhibit a 100\% binary fraction. They appear to be push-out survivors, emplaced in the classical region during Neptune's phases of outward migration. The presence of these binary systems implies that the majority of objects that formed near the cold classical region formed as binaries.  Here we present new optical colour measurements of cold classical KBOs from the Colours of the Outer Solar System Origins Survey, including colours of a blue binary discovered by the Solar System Origins Legacy Survey -- 2015 RJ277. The increased size of the colours sample has resulted in order-of-magnitude decrease  in the probability that the binaries and singles sample share the same colour distribution. From the Anderson-Darling statistic, this probability is only a 0.3\%, while it is only 0.002\% when utilizing the difference of means statistic. We find a hint that the blue binaries have inflated free inclinations compared to their red counterparts, consistent with the push-out origin for these bodies.
\end{abstract}


\section{Introduction \label{sec:Intro}}

The low inclination component of the classical Kuiper Belt is thought to be the most dynamically pristine planetesimal population in the trans-Neptunian region. In particular, the low-i objects with semi-major axes between the 3:2 and 2:1 mean-motion resonances with Neptune, often referred to as the \coldclassical~ Kuiper Belt Objects (CCKBOs) are dynamically isolated from Neptune, and are frequently found in dynamically fragile, widely separated binary pairs \citep{Petit2004,Parker2012}. It is for these two reasons that the CCKBOs are thought to be the only population of trans-Neptunian Objects that formed mainly in-situ, rather than having been emplaced on their current orbits through scattering by Neptune \citep{Parker2010a,Nesvorny2019b}, as many scattering and resonant objects have. 

While the majority of objects within the CCKBO population are distinctly red, there is a growing sample of bluer objects in the CCKBO region.  The blue binary CCKBOs have been identified via their distinct surface colours \citep{Fraser2017}. Named for their surfaces with bluer spectral slopes than found for most CCKBOs \citep[eg.][]{Gulbis2006, Doressoundiram2007} the so-called \bluebinaries~ have optical colours that overlap those of the less red \citep[LR;][]{Wong2017, Schwamb2019} compositional class of all trans-Neptunian Objects (TNOs). Those colours are unlike all other known CCKBOs that all have colours consistent with the very-red (VR) class \citep{Fraser2012, Peixinho2015}. 

Based on early results from the Colours of the Outer Solar System Origins Survey (Col-OSSOS), in \citet{Fraser2017} we hypothesized that the \bluebinaries~ are contaminants to the cold classical region, having been pushed out from a region only slightly interior to the current location of the 3:2 resonance during a phase of Neptune's \emph {smooth} outward migration in one of the strong mean-motion resonances. The hypothesis that the \bluebinaries~ are push-out survivors comes with the startling expectation that the majority of objects in the cold classical region formed in binary pairs \citep{Fraser2017}. Evidence for this has been building since the discovery of wide binaries. From a dynamical analysis of objects like 2001 QW322 - now recognized as a blue binary - \citet{Petit2004} pointed out that in the early Solar System, binaries must have been an order of magnitude more numerous than currently observed, though only 10-30\% of CCKBOs are resolved binary systems \citep{Noll2020tnss}. One formation route in which such a scenario is possible is if the binary KBOs are the products of the gravitational collapse of a self-gravitating pebble cloud. During gravitational collapse, due to the conservation of angular momentum, pebble clouds will preferentially produce bound planetesimal systems of high multiplicity with masses and orbital properties comparable to observed KBO binary systems \citep{Nesvorny2010, Parker2011,  Nesvorny2019a, Grundy2019, Robinson2020}. 

The existence of the \bluebinaries~ as objects pushed out to their current locations indicates that the conditions needed to produce a high binary fraction must have extended into the $\sim30-40$~AU zone of the primordial Solar System. The push-out origin theory for the \bluebinaries~ hinges on the idea that the binary CCKBOs possess a different surface colour distribution than do the single (or unresolved) CCKBOs. If the single and binary CCKBOs do not possess distinct colour distributions, then there is no evidence to suggest that the binary CCKBOs possess a population of LR objects that is not present in the singles. It is because the observed binary CCKBO colour distribution extends to bluer surface colours than that observed for the singles, that the presence of the \bluebinaries~ has been inferred (see Figure~\ref{fig:slopes}). 

Here we increase the sample of CCKBO optical colours, and consider recent identifications of binary KBOs made with high spatial resolution observations. Specifically, we make use of observations from the Colours of the Outer Solar System Origins Survey \citep[Col-OSSOS;][]{Schwamb2019} and the Solar System Origins Legacy Survey \citep[SSOLS;][]{Parker2020DPS, Benecchi2020DPS}. This significantly increases the sample of binary CCKBOs with which to test whether or not the binary colour distribution differs from that of the singles. Here we present a  more statistically robust demonstration that the binary CCKBOs have a colour distribution that is distinct from the single bodies.

In Section 2 we discuss the increased sample of colours and high resolution observations. In Section 3 we consider the orbital elements of our sample of CCKBOs with colour measurements, and present the possibility that the blue binaries on average, have higher free inclinations (the orbital inclinations  measured with respect to the forced orbital plane \citep{vanLaerhoven2019}) than their red counterparts.

\section{The Colours and Binaries Sample}
We adopt the usual definition of \coldclassical~ \citep[eg.][]{Elliot2005,Gladman2008}, with the usual orbital element selection typically used to select these objects: non-resonant objects with inclinations $i<6^\circ$, perihelion distances $q>37$~AU, and semi-major axes $42<a<47.5$~AU, the same definition we adopted in \citet{Fraser2017}. We also include objects in the 7:4 mean-motion resonance with Neptune, with inclinations $i<6^\circ$. The inclusion of these objects was made because low-i objects in this region can be temporarily resonant. In addition, most of these low-i resonators likely share a common origin with non-resonant CCKBOs. This is highlighted by the blue binary, 2016 BP81, which has been re-classed from non-resonant \citep{Bannister2016} to resonant \citep{Bannister2018}. This system  appears like all other blue binaries, and differs only in its transient resonant behaviour, due to its chance location near the 7:4, as demonstrated by numerical integration of its heliocentric orbit. We emphasize that even with the exclusion of all resonant and non-resonant objects within $\sim0.3$~AU of the 7:4 resonance we would still draw the same conclusions as presented below.

The aim of Col-OSSOS is to gather a brightness complete, UV-optical-NIR colours sample of objects discovered in certain blocks of the Outer Solar System Origins Survey \citep{Bannister2018} which has been gathering UV-optical-NIR colours with the Frederick C. Gillett Gemini North Telescope, and the Canada-France-Hawaii Telescope.  Details of Col-OSSOS observations, target selection, and data reductions can be found in \citet{Schwamb2019}. Here we report newly measured optical colours for 17 CCKBOs along with the complete sample of all CCKBOs measured by Col-OSSOS.

We further consider the binary classifications of the SSOLS project \citep{Parker2020DPS, Benecchi2020DPS}\footnote{\url{http://www2.lowell.edu/users/grundy/tnbs/status.html}}. The SSOLS program is conducting a search for binaries among the complete set of  OSSOS and the Canada-France Ecliptic Plane Survey (CFEPS) bright CCKBOs with aim to create a binary sample with well characterized discovery biases. Together, the new colour measurements and binary detections are used to supplement the sample we utilized in \citet{Fraser2017}.

We adopt the spectral slopes and methods in determining those slopes that were published in \citet{Fraser2017}. Specifically, the {\it synphot} tool was used to forward model a Solar spectrum convolved with a linear spectrum to estimate colours in appropriate bands as a function of spectral slope. Reported measurements are the weighted means of all available measures for a given target, and we only consider those mean spectral slope values with uncertainty $\Delta s<7\%/100\mbox{ $n$m}$. As the phase curves of virtually all objects in our sample are unknown, no effort to correct to $0^\circ$ phase angle is made.  Though we highlight the near-simultaneous nature of the observations from which the colours were measured, mitigating the need to account for differences in phase angle. The total sample of high quality CCKBO colours is increased from 87 to 113 (a 30\% increase), 30 of which have been identified as binaries. We present the full sample of spectral slopes, along with (g-r) colours for the Col-OSSOS targets in Table~\ref{tab:colours}.

\begin{figure}[h]
   \centering
   \plotone{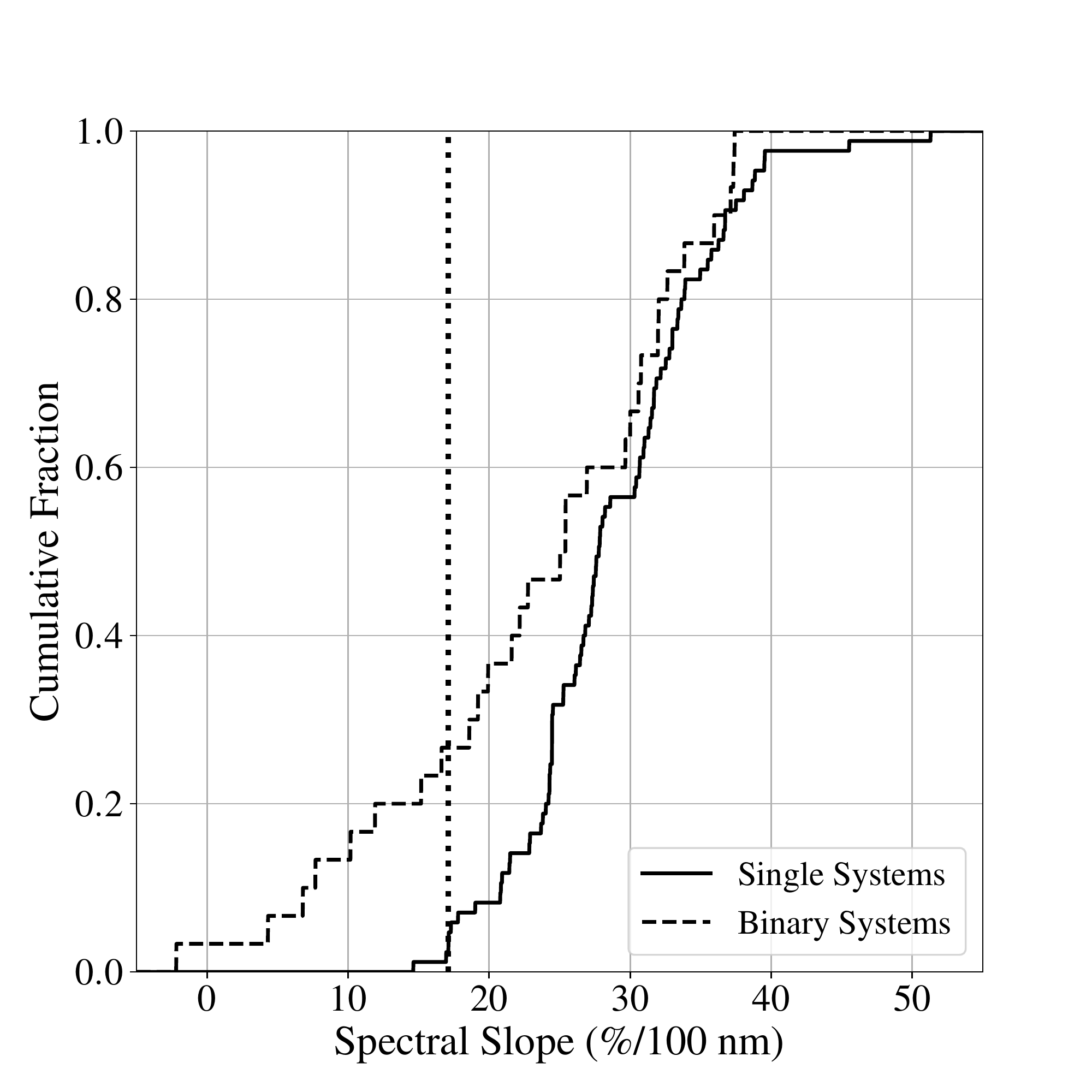} 
   \figcaption{Cumulative spectral slope distributions of the single (solid) and binary (dashed) samples. The vertical line at $s=17 \%/100\mbox{ nm}$ corresponds to (g-r)=0.75, and demarks the blue edge of the very red compositional class, as determined from the optical-NIR colour distribution of the dynamically excited TNOs \citep{Fraser2012, Schwamb2019}. \label{fig:slopes}}
\end{figure}

We consider two samples: those objects reported as binary, and the sample of singles, noting that the latter is likely made up of a combination of truly single bodies, and those with unresolved companions. We present the cumulative spectral slope histograms of the binary and single populations in Figure~\ref{fig:slopes}. 

2015 RJ277 stands out as the only new CCKBO identified with a slightly redder than Solar optical colour since our 2017 manuscript;  its spectral slope is $s=11.9\pm1.3 \%/100\mbox{ $n$m}$ consistent with the colours of the LR class. RJ277 is the 6th bluest surface in our sample (see Table 1). Hubble Space Telescope observations of this target from SSOLS have revealed this to be a binary system with an apparent separation of $\sim0.04"$. This system is indicative of the standout nature of the blue binaries: the 6 bluest objects in our sample of \coldclassicals~ are all binaries. Alternatively stated, 100\% of objects bluer than the bluest single object in our colours sample are binary.

\begin{figure}[h]
   \centering
   \plotone{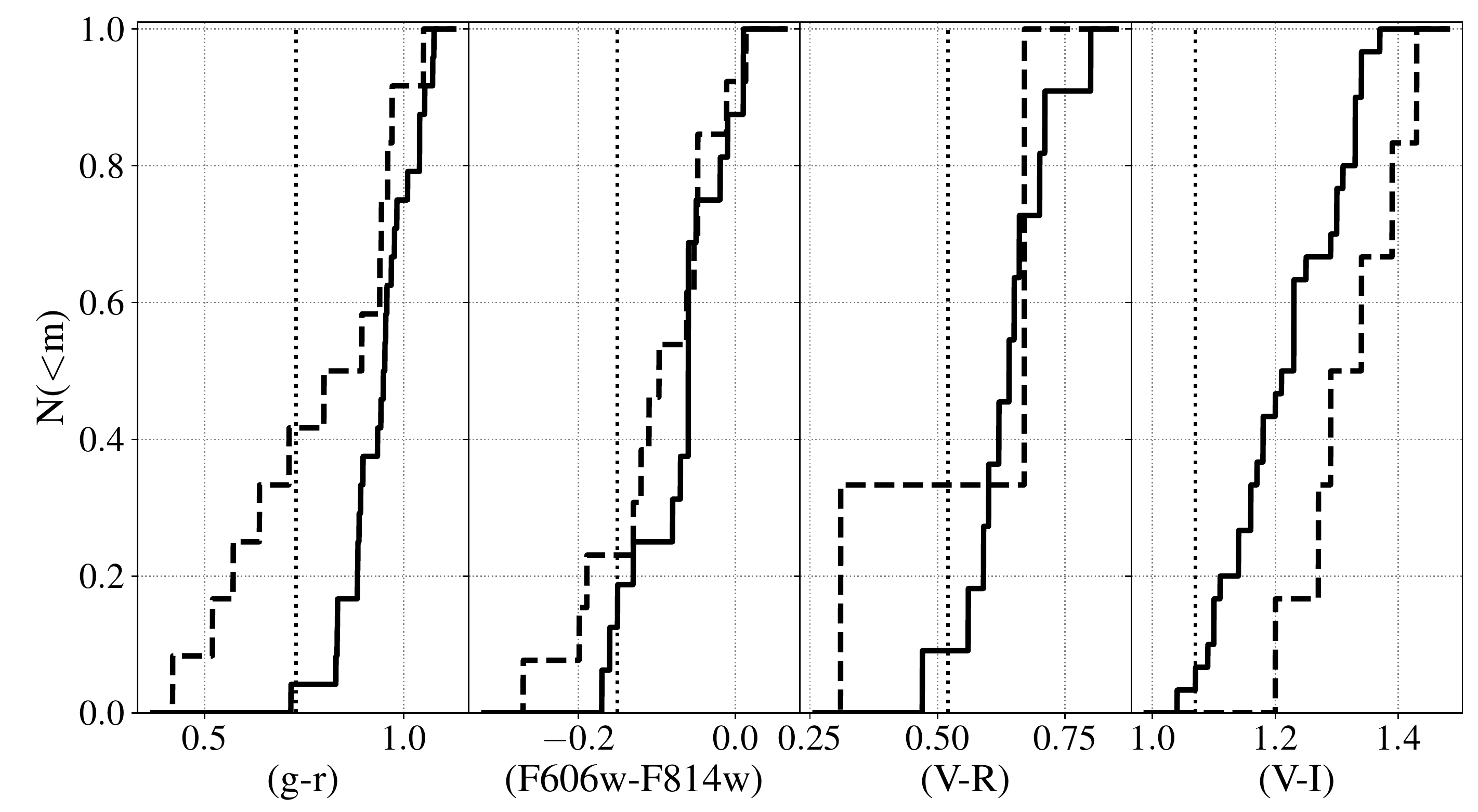} 
   \figcaption{Cumulative colour distributions of the single (solid) and binary (dashed) samples, that were used to evaluate mean spectral slopes $s$ presented in Figure~1. Like in Figure~1, we only plot those objects with spectral slope error less than $7\%/100\mbox{ nm}$. The vertical line shows the colour corresponding to $s=17\%/100\mbox{ $n$m}$. The bluest objects are binaries in all but (V-I) which happens to not include any of the blue binary CCKBOs.  \label{fig:colours}}
\end{figure}

It is apparent from Figure~\ref{fig:slopes} that the binary sample extends to lower spectral slopes  $s\sim0 \%/100\mbox{nm}$ than does the singles sample with the bluest single having spectral slope $s=14 \%/100\mbox{nm}$; the 6 objects with lowest spectral slope are binary. This feature of the binary colours sample is also apparent in the raw colours from which the spectral slopes were estimated (see Figure~\ref{fig:colours}). The bluest objects in (g-r), (V-R), and (F606w-F814w) are binary. The only colour for which this is not true is (V-I). By chance, the (V-I) sample does not include any of the \bluebinaries. It is worth noting that the majority of \bluebinaries~ arise from the Col-OSSOS sample. As the spectra of small KBOs are quite linear \citep[eg.][]{Barucci2011,Peixinho2015}, this feature of the colours sample is likely nothing more than a sampling fluke and is not indicative of different spectral behaviour in those bands.

Oddly, the singles sample also extends to redder spectral slopes; the 7 reddest targets in our full colours sample are singles. This is likely just an effect of the more populous singles sample, and probably does not reflect anything unusual about the red side of the colour distributions. If samples of size equal to the binary and single samples are drawn from the full colours sample, the probability that  the $\geq7$ reddest source belong to the singles population is 7\%.

We make use of two statistics to determine the probability that the single and binary samples share the same parent spectral slope distribution. As we did in \citet{Fraser2017}, we utilize the difference of means. We also make use of two-sample Anderson-Darling (AD) statistic \citep{Anderson1952,Efron1979,Scholz1987}. To determine the significance of either statistic, we implement bootstrapping with repeats and include additional scattering that matching the uncertainties. That is, from the single colours sample, a random spectral slope sample of size equal to the observed binary sample was bootstrapped, with each sampled point scattered by a gaussian. The widths of each gaussian were also bootstrapped from the distribution of spectral slope uncertainties of the single colours sample. The AD statistic using this random sample was then calculated. This process was repeated $10^6$ times, and the simulated statistic values were compared against the statistic evaluated from the observed binary and singles samples. The probability of drawing a simulated AD statistic equal to or larger than the real value was 0.3\%. That is to say, the probability that the single and binary samples share the same parent colour distribution is only 0.3\%. For the difference of means statistic, the probability of the null hypothesis was only 0.002\%. The difference in results from the two statistics reflects their relative sensitivities to differences in the tails of two cumulative distributions, which is exactly where the single and binary distributions differ the most. More compelling than the statistics, however, is the nature of our result.  In \citet{Fraser2017} we predicted that the rare blue members of the CCKBO group are all binary. The only new blue member added in the last 4 years is, indeed, binary, consistent with our earlier prediction.

\section{Considering Orbital Elements}

\begin{figure*}[h]
   \centering
   \plotone{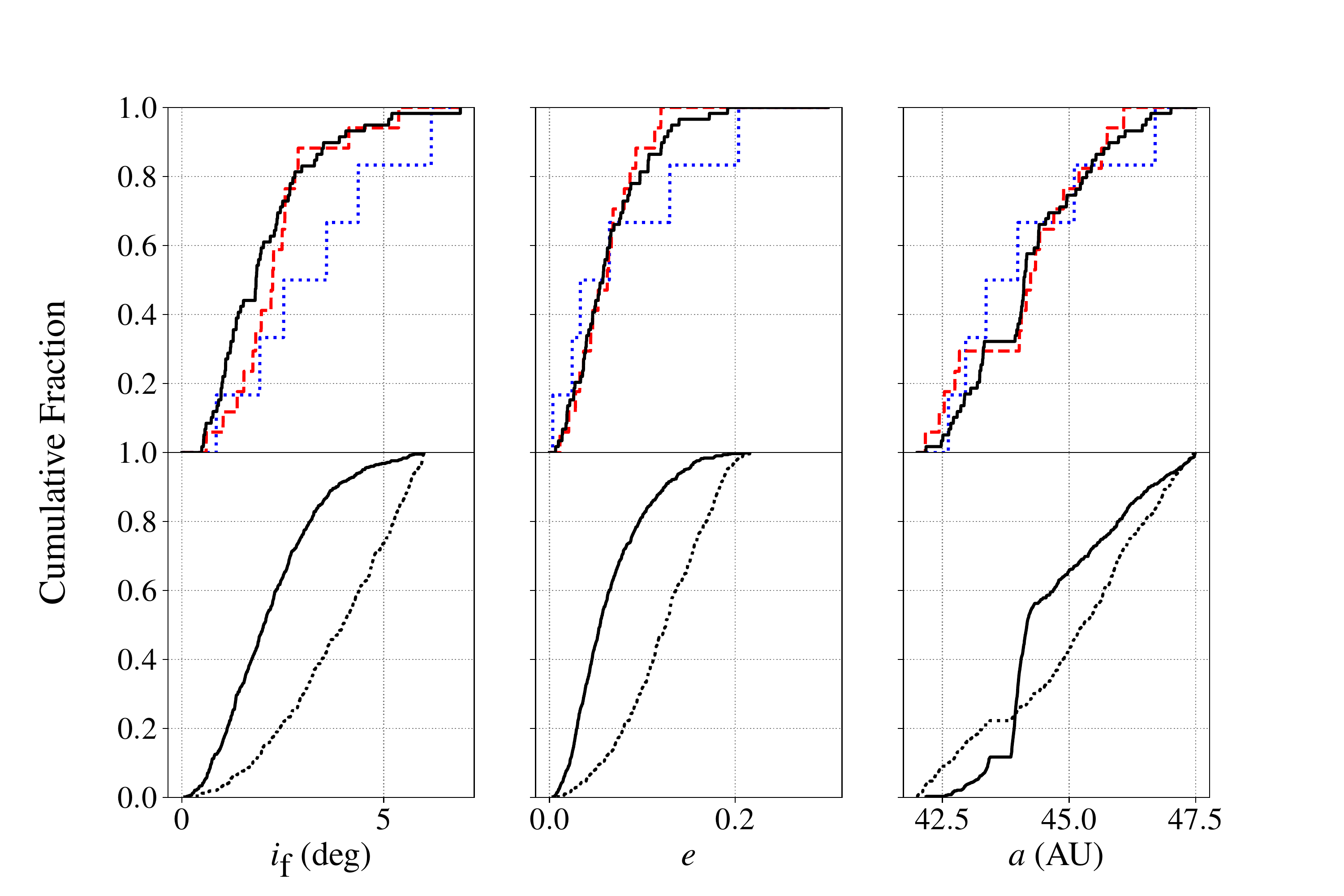} 
   \figcaption{Cumulative free inclination, eccentricity, and semi-major axis of the TNO colours sample (top), and from the results of a migration simulation from \citet{Nesvorny2015a} (bottom). The element distributions of the VR binaries, LR binaries, and single objects are shown by the red, blue, and black curves, respectively.  From the simulation, the dashed and solid lines present the element distributions of those objects ending as cold classicals, that originated in the cold classical region (initial $a>42$~AU; solid curve), and those originating interior to the region  (initial $30<a<41$~AU; dashed curve). We emphasize that a different selection of initial distances for those objects originating interior to the cold classical region makes no appreciable difference on the final orbital element distributions. We do not include the small contribution from objects originating inside $30$~AU, as binarity would be disrupted from push-out over such large distances. The overall contribution of these bodies to the cold classical population will be considered in a future work.  Free inclinations of real CCKBOs are from \citep{vanLaerhoven2019}. \label{fig:freeincs}}
\end{figure*}

If  \bluebinaries~ are push-out survivors, it may be that they possess a different orbital element distribution than do the single objects, or even the very red (VR) binaries. The idea that the \bluebinaries~ have been pushed out during Neptune's migration suggests that on average, the \bluebinaries~ should possess higher free inclinations than do the red CCKBOs, as generally the push-out process acts to inflate an object's inclination \citep{Malhotra1993,Hahn2005,Nesvorny2015b}. This is shown in Figure~\ref{fig:freeincs}, where we plot the element distributions of an example migration simulation from \citet{Nesvorny2015a}. This simulation is typical of those that provide a decent match to the observed elements, and uses grainy migration with timescale of $\tau=10$~Myr, and a disk with exponential profile with eccentricity folding value of $3$~AU. From this simulation, it is clear that push-out objects are preferentially found on more excited orbits, with higher eccentricities and inclinations than their non--push-out counterparts \citep[also, see][]{Fraser2017}.

\begin{figure}[h]
   \centering
   \plotone{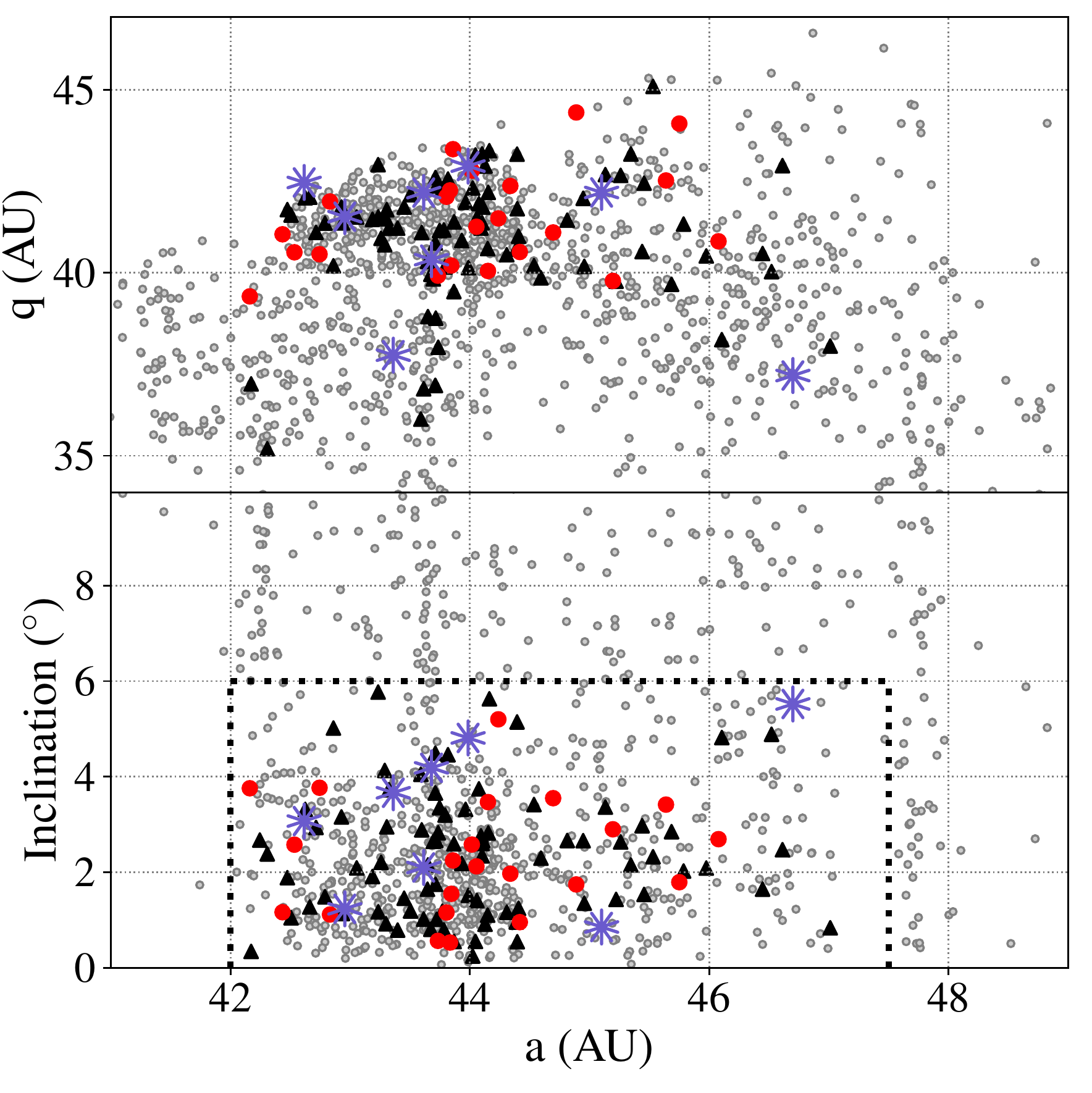} 
   \figcaption{Perihelion distance and inclination versus semi-major axis of the LR (blue stars) and VR (red circles) binaries and single objects (black triangles) in our colours sample. The MPC tabulated objects are shown for reference. The dashed goal posts show the cuts in inclination and semi-major axis made to select cold classicals. \label{fig:elems}}
\end{figure}

We consider the barycentric orbital elements of our sample of CCKBO with measured colours. We present their orbital elements, cumulatively in Figure~\ref{fig:freeincs}, and individually in Figure~\ref{fig:elems}. Generally, with the AD statistic, we have found no statistically significant differences in the orbital element distributions between any of the binary samples or that of the singles. The only hint of a difference in orbital elements between the two samples arises from their free inclinations, $i_\textrm{f}$. The values of $i_\textrm{f}$, are the orbital inclinations  measured with respect to the forced orbital plane \citep{vanLaerhoven2019}.  When considering $i_\textrm{f}$, we have removed all objects with semi-major axes, $43.45<a<43.65$~AU as they may interact with the 7:4 resonance, which would invalidate our calculation of $i_\textrm{f}$. From the cumulative distributions, it seems that the singles and red binaries are preferentially found on orbits with low-$i_\textrm{f}$, with 80\% of those objects found with $i_\textrm{f}\lesssim3^\circ$. The \bluebinaries~ however, appear to be more commonly found on higher inclination orbits, with half possessing inclinations $i_\textrm{f}>3^\circ$. This distinction is not significant however, and is only indicative of what signals might be searched for as sample sizes grow.

The colours sample appears to show a similar preference of higher $i_{\textrm{f}}$ for the blue binaries, compatible with the idea that these objects are push-out survivors. By definition of the cold classical object, the inclination inflation experienced by the \bluebinaries~ must be small, which may explain why the difference in cumulative $i_\textrm{f}$ distributions between the LR binaries and the CCKBOs in general is not particularly large and is not statistically significant. We applied the AD statistic in the same manner as for the spectral slope distribution and found a 20\% chance that the the $i_\textrm{f}$ distributions share the same parent distribution. 

Increasing the sample size of LR CCKBOs with surveys that involve optical colour measurements, and high resolution observations, is a worthwhile pursuit. If it is shown that the $i_\textrm{f}$ (or $e$) distribution of the LR binaries is different from the VR CCKBOs, it would provide an important lever with which to constrain their initial inclination distributions, and the push-out distance experienced by those bodies. Moreover, the orbital element distributions of the blue binaries have the potential to reveal details of the migratory path experienced by Neptune during push-out of those objects.

\acknowledgements

The authors acknowledge the sacred nature of Maunakea and appreciate the opportunity to observe from the mountain. This work is based on observations from various programs at the Gemini Observatory, which is operated by the Association of Universities for Research in Astronomy, Inc., under a cooperative agreement with the NSF on behalf of the Gemini partnership: the National Science Foundation (United States), the National Research Council (Canada), CONICYT (Chile), Ministerio de Ciencia, Tecnolog\'{i}a e Innovaci\'{o}n Productiva (Argentina), and Minist\'{e}rio da Ci\^{e}ncia, Tecnologia e Inova\c{c}\~{a}o (Brazil). This work is also based on observations obtained with MegaPrime/MegaCam, a joint project of CFHT and CEA/DAPNIA, at the Canada–France–Hawaii Telescope (CFHT), which is operated by the National Research Council (NRC) of Canada, the Institut National des Sciences de l’Univers of the Centre National de la Recherche Scientifique of France, and the University of Hawaii.
Benecchi acknowledges support from HST program GO-15648 provided from NASA through a grant from the Space Telescope Science Institute, which is operated by the Association of Universities for Research in Astronomy, Inc., under NASA contract NAS 5-26555.

\bibliographystyle{aasjournal}
\bibliography{astroelsart}

\begin{thebibliography}{}
\expandafter\ifx\csname natexlab\endcsname\relax\def\natexlab#1{#1}\fi
\providecommand{\url}[1]{\href{#1}{#1}}
\providecommand{\dodoi}[1]{doi:~\href{http://doi.org/#1}{\nolinkurl{#1}}}
\providecommand{\doeprint}[1]{\href{http://ascl.net/#1}{\nolinkurl{http://ascl.net/#1}}}
\providecommand{\doarXiv}[1]{\href{https://arxiv.org/abs/#1}{\nolinkurl{https://arxiv.org/abs/#1}}}

\bibitem[{Anderson \& Darling(1952)}]{Anderson1952}
Anderson, T.~W., \& Darling, D.~A. 1952, The Annals of Mathematical Statistics,
  23, 193.
\newblock \url{http://www.jstor.org/stable/2236446}

\bibitem[{{Bannister} {et~al.}(2016){Bannister}, {Kavelaars}, {Petit},
  {Gladman}, {Gwyn}, {Chen}, {Volk}, {Alexandersen}, {Benecchi}, {Delsanti},
  {Fraser}, {Granvik}, {Grundy}, {Guilbert-Lepoutre}, {Hestroffer}, {Ip},
  {Jakubik}, {Jones}, {Kaib}, {Kavelaars}, {Lacerda}, {Lawler}, {Lehner},
  {Lin}, {Lister}, {Lykawka}, {Monty}, {Marsset}, {Murray-Clay}, {Noll},
  {Parker}, {Pike}, {Rousselot}, {Rusk}, {Schwamb}, {Shankman}, {Sicardy},
  {Vernazza}, \& {Wang}}]{Bannister2016}
{Bannister}, M.~T., {Kavelaars}, J.~J., {Petit}, J.-M., {et~al.} 2016, \aj,
  152, 70, \dodoi{10.3847/0004-6256/152/3/70}

\bibitem[{{Bannister} {et~al.}(2018){Bannister}, {Gladman}, {Kavelaars},
  {Petit}, {Volk}, {Chen}, {Alexand ersen}, {Gwyn}, {Schwamb}, {Ashton},
  {Benecchi}, {Cabral}, {Dawson}, {Delsanti}, {Fraser}, {Granvik},
  {Greenstreet}, {Guilbert-Lepoutre}, {Ip}, {Jakubik}, {Jones}, {Kaib},
  {Lacerda}, {Van Laerhoven}, {Lawler}, {Lehner}, {Lin}, {Lykawka}, {Marsset},
  {Murray-Clay}, {Pike}, {Rousselot}, {Shankman}, {Thirouin}, {Vernazza}, \&
  {Wang}}]{Bannister2018}
{Bannister}, M.~T., {Gladman}, B.~J., {Kavelaars}, J.~J., {et~al.} 2018, \apjs,
  236, 18, \dodoi{10.3847/1538-4365/aab77a}

\bibitem[{{Barucci} {et~al.}(2011){Barucci}, {Alvarez-Candal}, {Merlin},
  {Belskaya}, {de Bergh}, {Perna}, {DeMeo}, \& {Fornasier}}]{Barucci2011}
{Barucci}, M.~A., {Alvarez-Candal}, A., {Merlin}, F., {et~al.} 2011, \icarus,
  214, 297, \dodoi{10.1016/j.icarus.2011.04.019}

\bibitem[{{Benecchi} {et~al.}(2020){Benecchi}, {Parker}, {Porter}, {Noll},
  {Grundy}, {Bannister}, \& {Kavelaars}}]{Benecchi2020DPS}
{Benecchi}, S., {Parker}, A., {Porter}, S., {et~al.} 2020, in AAS/Division for
  Planetary Sciences Meeting Abstracts, Vol.~52, AAS/Division for Planetary
  Sciences Meeting Abstracts, 307.02

\bibitem[{{Doressoundiram} {et~al.}(2007){Doressoundiram}, {Peixinho},
  {Moullet}, {Fornasier}, {Barucci}, {Beuzit}, \&
  {Veillet}}]{Doressoundiram2007}
{Doressoundiram}, A., {Peixinho}, N., {Moullet}, A., {et~al.} 2007, \aj, 134,
  2186, \dodoi{10.1086/522783}

\bibitem[{Efron \& Stein(1981)}]{Efron1979}
Efron, B., \& Stein, C. 1981, The Annals of Statistics, 9, 586.
\newblock \url{http://www.jstor.org/stable/2240822}

\bibitem[{{Elliot} {et~al.}(2005){Elliot}, {Kern}, {Clancy}, {Gulbis},
  {Millis}, {Buie}, {Wasserman}, {Chiang}, {Jordan}, {Trilling}, \&
  {Meech}}]{Elliot2005}
{Elliot}, J.~L., {Kern}, S.~D., {Clancy}, K.~B., {et~al.} 2005, AJ, 129, 1117,
  \dodoi{10.1086/427395}

\bibitem[{{Fraser} \& {Brown}(2012)}]{Fraser2012}
{Fraser}, W.~C., \& {Brown}, M.~E. 2012, \apj, 749, 33,
  \dodoi{10.1088/0004-637X/749/1/33}

\bibitem[{{Fraser} {et~al.}(2017){Fraser}, {Bannister}, {Pike}, {Marsset},
  {Schwamb}, {Kavelaars}, {Lacerda}, {Nesvorn{\'y}}, {Volk}, {Delsanti},
  {Benecchi}, {Lehner}, {Noll}, {Gladman}, {Petit}, {Gwyn}, {Chen}, {Wang},
  {Alexandersen}, {Burdullis}, {Sheppard}, \& {Trujillo}}]{Fraser2017}
{Fraser}, W.~C., {Bannister}, M.~T., {Pike}, R.~E., {et~al.} 2017, Nature
  Astronomy, 1, 0088, \dodoi{10.1038/s41550-017-0088}

\bibitem[{{Gladman} {et~al.}(2008){Gladman}, {Marsden}, \&
  {Vanlaerhoven}}]{Gladman2008}
{Gladman}, B., {Marsden}, B.~G., \& {Vanlaerhoven}, C. 2008, {Nomenclature in
  the Outer Solar System} (The Solar System Beyond Neptune), 43--57

\bibitem[{{Grundy} {et~al.}(2019){Grundy}, {Noll}, {Roe}, {Buie}, {Porter},
  {Parker}, {Nesvorn{\'y}}, {Levison}, {Benecchi}, {Stephens}, \&
  {Trujillo}}]{Grundy2019}
{Grundy}, W.~M., {Noll}, K.~S., {Roe}, H.~G., {et~al.} 2019, \icarus, 334, 62,
  \dodoi{10.1016/j.icarus.2019.03.035}

\bibitem[{{Gulbis} {et~al.}(2006){Gulbis}, {Elliot}, \& {Kane}}]{Gulbis2006}
{Gulbis}, A. A.~S., {Elliot}, J.~L., \& {Kane}, J.~F. 2006, \icarus, 183, 168,
  \dodoi{10.1016/j.icarus.2006.01.021}

\bibitem[{{Hahn} \& {Malhotra}(2005)}]{Hahn2005}
{Hahn}, J.~M., \& {Malhotra}, R. 2005, \aj, 130, 2392, \dodoi{10.1086/452638}

\bibitem[{{Malhotra}(1993)}]{Malhotra1993}
{Malhotra}, R. 1993, \nat, 365, 819, \dodoi{10.1038/365819a0}

\bibitem[{{Marsset} {et~al.}(2020){Marsset}, {Fraser}, {Bannister}, {Schwamb},
  {Pike}, {Benecchi}, {Kavelaars}, {Alexandersen}, {Chen}, {Gladman}, {Gwyn},
  {Petit}, \& {Volk}}]{Marsset2020}
{Marsset}, M., {Fraser}, W.~C., {Bannister}, M.~T., {et~al.} 2020, The
  Planetary Science Journal, 1, 16, \dodoi{10.3847/PSJ/ab8cc0}

\bibitem[{{Nesvorn{\'y}}(2015{\natexlab{a}})}]{Nesvorny2015b}
{Nesvorn{\'y}}, D. 2015{\natexlab{a}}, \aj, 150, 73,
  \dodoi{10.1088/0004-6256/150/3/73}

\bibitem[{{Nesvorn{\'y}}(2015{\natexlab{b}})}]{Nesvorny2015a}
---. 2015{\natexlab{b}}, \aj, 150, 68, \dodoi{10.1088/0004-6256/150/3/68}

\bibitem[{{Nesvorn{\'y}} {et~al.}(2019){Nesvorn{\'y}}, {Li}, {Youdin}, {Simon},
  \& {Grundy}}]{Nesvorny2019a}
{Nesvorn{\'y}}, D., {Li}, R., {Youdin}, A.~N., {Simon}, J.~B., \& {Grundy},
  W.~M. 2019, Nature Astronomy, 3, 808, \dodoi{10.1038/s41550-019-0806-z}

\bibitem[{{Nesvorn{\'y}} \& {Vokrouhlick{\'y}}(2019)}]{Nesvorny2019b}
{Nesvorn{\'y}}, D., \& {Vokrouhlick{\'y}}, D. 2019, \icarus, 331, 49,
  \dodoi{10.1016/j.icarus.2019.04.030}

\bibitem[{{Nesvorn{\'y}} {et~al.}(2010){Nesvorn{\'y}}, {Youdin}, \&
  {Richardson}}]{Nesvorny2010}
{Nesvorn{\'y}}, D., {Youdin}, A.~N., \& {Richardson}, D.~C. 2010, \aj, 140,
  785, \dodoi{10.1088/0004-6256/140/3/785}

\bibitem[{{Noll} {et~al.}(2020){Noll}, {Grundy}, {Nesvorn{\'y}}, \&
  {Thirouin}}]{Noll2020tnss}
{Noll}, K., {Grundy}, W.~M., {Nesvorn{\'y}}, D., \& {Thirouin}, A. 2020,
  {Trans-Neptunian binaries (2018)}, ed. D.~{Prialnik}, M.~A. {Barucci}, \&
  L.~{Young}, 201--224, \dodoi{10.1016/B978-0-12-816490-7.00009-6}

\bibitem[{{Parker} {et~al.}(2020){Parker}, {Benecchi}, {Grundy}, {Kavelaars},
  {Bannister}, {Noll}, \& {Porter}}]{Parker2020DPS}
{Parker}, A., {Benecchi}, S., {Grundy}, W., {et~al.} 2020, in AAS/Division for
  Planetary Sciences Meeting Abstracts, Vol.~52, AAS/Division for Planetary
  Sciences Meeting Abstracts, 307.01

\bibitem[{{Parker} \& {Kavelaars}(2010)}]{Parker2010a}
{Parker}, A.~H., \& {Kavelaars}, J.~J. 2010, \apjl, 722, L204,
  \dodoi{10.1088/2041-8205/722/2/L204}

\bibitem[{{Parker} \& {Kavelaars}(2012)}]{Parker2012}
---. 2012, \apj, 744, 139, \dodoi{10.1088/0004-637X/744/2/139}

\bibitem[{{Parker} {et~al.}(2011){Parker}, {Kavelaars}, {Petit}, {Jones},
  {Gladman}, \& {Parker}}]{Parker2011}
{Parker}, A.~H., {Kavelaars}, J.~J., {Petit}, J.-M., {et~al.} 2011, \apj, 743,
  1, \dodoi{10.1088/0004-637X/743/1/1}

\bibitem[{{Peixinho} {et~al.}(2015){Peixinho}, {Delsanti}, \&
  {Doressoundiram}}]{Peixinho2015}
{Peixinho}, N., {Delsanti}, A., \& {Doressoundiram}, A. 2015, \aap, 577, A35,
  \dodoi{10.1051/0004-6361/201425436}

\bibitem[{{Petit} \& {Mousis}(2004)}]{Petit2004}
{Petit}, J.~M., \& {Mousis}, O. 2004, \icarus, 168, 409,
  \dodoi{10.1016/j.icarus.2003.12.013}

\bibitem[{{Pike} {et~al.}(2017){Pike}, {Fraser}, {Schwamb}, {Kavelaars},
  {Marsset}, {Bannister}, {Lehner}, {Wang}, {Alexandersen}, {Chen}, {Gladman},
  {Gwyn}, {Petit}, \& {Volk}}]{Pike2017}
{Pike}, R.~E., {Fraser}, W.~C., {Schwamb}, M.~E., {et~al.} 2017, \aj, 154, 101,
  \dodoi{10.3847/1538-3881/aa83b1}

\bibitem[{{Robinson} {et~al.}(2020){Robinson}, {Fraser}, {Fitzsimmons}, \&
  {Lacerda}}]{Robinson2020}
{Robinson}, J.~E., {Fraser}, W.~C., {Fitzsimmons}, A., \& {Lacerda}, P. 2020,
  \aap, 643, A55, \dodoi{10.1051/0004-6361/202037456}

\bibitem[{Scholz \& Stephens(1987)}]{Scholz1987}
Scholz, F.~W., \& Stephens, M.~A. 1987, Journal of the American Statistical
  Association, 82, 918, \dodoi{10.1080/01621459.1987.10478517}

\bibitem[{{Schwamb} {et~al.}(2019){Schwamb}, {Fraser}, {Bannister}, {Marsset},
  {Pike}, {Kavelaars}, {Benecchi}, {Lehner}, {Wang}, {Thirouin}, {Delsanti},
  {Peixinho}, {Volk}, {Alexandersen}, {Chen}, {Gladman}, {Gwyn}, \&
  {Petit}}]{Schwamb2019}
{Schwamb}, M.~E., {Fraser}, W.~C., {Bannister}, M.~T., {et~al.} 2019, \apjs,
  243, 12, \dodoi{10.3847/1538-4365/ab2194}

\bibitem[{{Van Laerhoven} {et~al.}(2019){Van Laerhoven}, {Gladman}, {Volk},
  {Kavelaars}, {Petit}, {Bannister}, {Alexandersen}, {Chen}, \&
  {Gwyn}}]{vanLaerhoven2019}
{Van Laerhoven}, C., {Gladman}, B., {Volk}, K., {et~al.} 2019, \aj, 158, 49,
  \dodoi{10.3847/1538-3881/ab24e1}

\bibitem[{{Wong} \& {Brown}(2017)}]{Wong2017}
{Wong}, I., \& {Brown}, M.~E. 2017, \aj, 153, 145,
  \dodoi{10.3847/1538-3881/aa60c3}

\end{thebibliography}

\clearpage

\startlongtable
\begin{deluxetable}{rlcc}
\tablenum{1}
\tablecaption{Colours of Cold Classical KBOs. \label{tab:colours}}
\tablewidth{0pt}
\tabletypesize{\footnotesize}
\tablehead{
\colhead{Target} & \colhead{(g-r)} & \colhead{$s ~(\%/100\mbox{ nm})$} & \colhead{Binary/Single}
}
\decimalcolnumbers
\startdata
1992 QB1 - Albion           &       -       & $23.8\pm2.3$ & S \\
1994 JQ1                    &       -       & $35.8\pm3.0$ & S \\
1994 VK8                    &       -       & $26.1\pm3.1$ & S \\
1995 DC2                    &       -       & $45.5\pm1.4$ & S \\
1995 WY2                    &       -       & $24.3\pm6.0$ & S \\
1996 TK66                   &       -       & $26.7\pm3.9$ & S \\
1997 CQ29 - Logos           &       -       & $18.6\pm5.0$ & B \\
1997 CS29 - Sila-Nunam      &       -       & $29.7\pm1.5$ & B \\
1997 CT29                   &       -       & $38.1\pm5.1$ & S \\
1997 CU29                   &       -       & $33.8\pm2.1$ & S \\
1998 HM151                  &       -       & $27.9\pm6.0$ & S \\
1998 HP151                  &       -       & $26.5\pm6.3$ & S \\
1998 KG62                   &       -       & $31.7\pm3.6$ & S \\
1998 KR65                   &       -       & $30.7\pm1.5$ & S \\
1998 KS65                   &       -       & $27.2\pm2.4$ & S \\
1998 WA25                   &       -       & $17.1\pm6.0$ & S \\
1998 WX31                   &       -       & $33.4\pm0.4$ & S \\
1998 WX24                   &       -       & $35.5\pm4.3$ & S \\
1998 WY24                   &       -       & $26.2\pm2.6$ & S \\
1999 CO153                  &       -       & $39.5\pm3.6$ & S \\
1999 HG12                   &       -       & $26.4\pm1.7$ & S \\
1999 HS11                   &       -       & $32.8\pm3.3$ & S \\
1999 HV11                   &       -       & $24.3\pm2.3$ & S \\
1999 OE4                    &       -       & $14.6\pm5.0$ & S \\
1999 OF4                    &       -       & $30.6\pm6.0$ & S \\
1999 OM4                    &       -       & $21.5\pm3.3$ & S \\
1999 RC215                  &       -       & $39.5\pm5.3$ & S \\
1999 RE215                  &       -       & $36.3\pm2.5$ & S \\
1999 RT214                  &       -       & $37.4\pm4.4$ & B \\
1999 RX214                  &       -       & $20.9\pm1.0$ & S \\
1999 RZ253 - Borasisi       &       -       & $37.1\pm4.7$ & B \\
2000 CE105                  &       -       & $28.6\pm4.3$ & S \\
2000 CF105                  &       -       & $21.6\pm6.8$ & B \\
2000 CH105                  &       -       & $26.8\pm1.2$ & S \\
2000 CL104                  &       -       & $21.4\pm2.4$ & S \\
2000 CM105                  &       -       & $33.8\pm7.0$ & B \\
2000 CN105                  &       -       & $31.4\pm3.8$ & S \\
2000 CQ114                  &       -       & $36.0\pm0.6$ & B \\
2000 FS53                   &       -       & $33.9\pm2.0$ & S \\
2000 OH67                   &       -       & $24.5\pm5.1$ & S \\
2000 OJ67                   &       -       & $26.9\pm3.0$ & B \\
2000 OK67                   &       -       & $19.0\pm1.9$ & S \\
2000 QC226                  &       -       & $51.3\pm5.6$ & S \\
2000 QN251                  &       -       & $24.5\pm0.2$ & S \\
2000 WK183                  &       -       & $25.4\pm2.6$ & B \\
2000 WT169                  &       -       & $19.9\pm2.0$ & B \\
2001 FK185*                  & $0.83\pm0.03$ & $24.2\pm2.0$ & S \\
2001 HZ58                   &       -       & $22.8\pm5.4$ & S \\
2001 KK76                   &       -       & $27.1\pm3.6$ & S \\
2001 KN76                   &       -       & $20.8\pm5.0$ & S \\
2001 KO76                   &       -       & $33.0\pm0.2$ & S \\
2001 OQ108                  &       -       & $30.3\pm4.4$ & S \\
2001 QE298                  & $0.83\pm0.02$ & $24.3\pm1.1$ & S \\
2001 QF331                  & $0.83\pm0.03$ & $24.0\pm1.5$ & S \\
2001 QO297                  &       -       & $32.2\pm3.7$ & S \\
2001 QP297                  &       -       & $27.6\pm1.2$ & S \\
2001 QR297                  &       -       & $24.5\pm4.0$ & S \\
2001 QS322                  &       -       & $22.9\pm3.3$ & S \\
2001 QT297 - Teharonhiawako &       -       & $25.4\pm2.3$ & B \\
2001 QW322                  &       -       & $-2.2\pm3.3$ & B \\
2001 QX297                  &       -       & $24.5\pm3.5$ & S \\
2001 QY297                  &       -       & $30.0\pm2.1$ & B \\
2001 RZ143                  &       -       & $19.2\pm2.9$ & B \\
2001 UQ18 - Altjira         &       -       & $32.0\pm4.1$ & B \\
2001 XR254                  &       -       & $15.2\pm4.4$ & B \\
2002 CC249                  &       -       & $20.8\pm4.6$ & S \\
2002 CU154                  &       -       & $24.5\pm4.0$ & S \\
2002 PA149                  &       -       & $25.3\pm2.5$ & S \\
2002 PD155                  &       -       & $17.8\pm5.6$ & S \\
2002 PV170                  &       -       & $23.7\pm2.2$ & S \\
2002 VD131                  &       -       & $ 6.8\pm3.6$ & B \\
2002 VT130                  &       -       & $22.7\pm1.6$ & B \\
2002 VV130                  &       -       & $36.6\pm0.9$ & S \\
2003 GH55                   &       -       & $24.5\pm1.8$ & S \\
2003 HG57                   &       -       & $10.2\pm3.4$ & B \\
2003 QF113                  &       -       & $25.3\pm3.4$ & S \\
2003 QW111 - Manwe          &       -       & $27.8\pm3.5$ & S \\
2003 QY111                  &       -       & $17.3\pm5.6$ & S \\
2003 QY90                   &       -       & $30.6\pm3.1$ & B \\
2003 TJ58                   &       -       & $25.0\pm2.2$ & B \\
2003 UN284                  &       -       & $ 4.3\pm5.8$ & B \\
2004 EU95*                   & $0.97\pm0.02$ & $32.5\pm1.4$ & S \\
2004 HJ79*                   & $0.95\pm0.02$ & $31.5\pm1.2$ & S \\
2004 PW107                  &       -       & $27.3\pm4.2$ & S \\
2005 EO304                  &       -       & $22.2\pm3.1$ & B \\
2006 BR284                  &       -       & $30.8\pm1.8$ & B \\
2006 CH69                   &       -       & $37.3\pm6.5$ & B \\
2006 HW122                  &       -       & $17.1\pm6.1$ & S \\
2006 JZ81                   &       -       & $32.0\pm4.2$ & B \\
2006 QF181                  & $0.90\pm0.03$ & $28.2\pm1.5$ & S \\
2013 EM149*                  & $0.96\pm0.02$ & $31.9\pm1.3$ & S \\
2013 GF138*                  & $1.07\pm0.03$ & $38.7\pm1.5$ & S \\
2013 GN137*                  & $1.05\pm0.01$ & $37.5\pm0.6$ & S \\
2013 GP137*                  & $0.94\pm0.03$ & $30.9\pm2.0$ & S \\
2013 GQ137*                  & $0.89\pm0.02$ & $27.8\pm1.3$ & S \\
2013 GR136*                  & $0.72\pm0.03$ & $16.9\pm1.6$ & S \\
2013 GS137*                  & $1.01\pm0.02$ & $35.0\pm1.3$ & S \\
2013 GT137*                  & $1.04\pm0.04$ & $36.7\pm2.2$ & S \\
2013 GV137*                  & $0.93\pm0.05$ & $30.4\pm2.7$ & S \\
2013 GW137*                  & $0.95\pm0.02$ & $31.3\pm1.4$ & S \\
2013 GX137*                  & $0.98\pm0.03$ & $33.4\pm1.7$ & S \\
2013 SP99                   & $0.98\pm0.02$ & $33.0\pm1.2$ & S \\
2013 SQ99*                   & $0.97\pm0.02$ & $32.6\pm1.4$ & B \\
2013 UL15                   & $0.90\pm0.03$ & $28.0\pm1.9$ & S \\
2013 UN15                   & $1.08\pm0.03$ & $38.8\pm1.9$ & S \\
2013 UO15                   & $0.95\pm0.02$ & $31.7\pm1.1$ & S \\
2013 UP15                   & $0.88\pm0.02$ & $27.4\pm1.2$ & S \\
2013 UX18*                   & $0.89\pm0.01$ & $27.6\pm0.6$ & S \\
2014 UD225                  & $0.71\pm0.02$ & $16.6\pm1.0$ & B \\
2014 UE225                  & $1.04\pm0.02$ & $36.7\pm1.0$ & S \\
2015 RB281*                  & $0.89\pm0.04$ & $27.4\pm2.4$ & S \\
2015 RJ277*                  & $0.64\pm0.02$ & $11.9\pm1.3$ & B \\
2015 RT245*                  & $0.94\pm0.05$ & $31.0\pm2.8$ & S \\
2016 BP81                   & $0.57\pm0.03$ & $ 7.7\pm1.7$ & B \\ 
\hline
2013 GV137* & $0.95\pm0.06$ & - & S \\ 
2013 GV137* & $0.92\pm0.02$ & - & S \\ 
2013 UN15 & $1.05\pm0.03$ & - & S \\ 
2013 UN15* & $1.09\pm0.04$ & - & S \\ 
2013 UN15* & $1.09\pm0.03$ & - & S \\ 
2015 RB281* & $0.77\pm0.04$ & - & S \\ 
2015 RB281* & $1.00\pm0.04$ & - & S \\ 
\enddata
\tablecomments{Col-OSSOS measurements are those reported with (g-r) colours. Measurements of targets reported in past Col-OSSOS papers have been updated to reflect improvements to the reductions pipelines since past publications \citep{Pike2017, Fraser2017, Schwamb2019, Marsset2020}, including improvements to PSF star selection and minor bug fixes. Spectral slope values were evaluated as in \citet{Fraser2017}. Targets marked with an asterisk are newly reported colours. The reported (g-r) colours of 2013 GV137, 2013 UN15, and 2015 RB281 are the mean of repeat measurements at each epoch, the values of which are at the bottom of the table. Target 2015 RB281 exhibits clear colour variability, which will be discussed in a future manuscript. Binary classifications are reported at \url{http://www2.lowell.edu/users/grundy/tnbs/}}
\end{deluxetable}

\clearpage

\end{document}